\documentclass[11pt]{article}
\usepackage{longtable,supertabular}
\usepackage{amsmath}
\usepackage{amssymb}
\usepackage{latexsym}
\usepackage{cite}

\newcommand{\tabincell}[2]{\begin{tabular}{@{}#1@{}}#2\end{tabular}}

\title{\bf New MDS Entanglement-Assisted Quantum Codes from  MDS Hermitian Self-Orthogonal Codes}
\author{Hao Chen
  \thanks{Hao Chen is with the College of Information Science and Technology/Cyber Security, Jinan University, Guangzhou, Guangdong Province, 510632, China, haochen@jnu.edu.cn. The research of Hao Chen was supported by NSFC Grant 62032009.}}

\begin{document}

\maketitle
\begin{abstract}
The intersection ${\bf C}\bigcap {\bf C}^{\perp_H}$ of a linear code ${\bf C} \subset {\bf F}_{q^2}$ and its Hermitian dual ${\bf C}^{\perp_H}$ is called the Hermitian hull of this code.  A linear code ${\bf C} \subset {\bf F}_{q^2}$ satisfying ${\bf C} \subset {\bf C}^{\perp_H}$ is called Hermitian self-orthogonal. Many Hermitian self-orthogonal codes were given for the construction of MDS quantum error correction codes (QECCs). In this paper we prove that for a nonnegative integer $h$ satisfying $0 \leq h \leq k$, a linear Hermitian self-orthogonal $[n, k]_{q^2}$ code is equivalent to a linear $h$-dimension Hermitian hull code. Therefore a lot of new MDS entanglement-assisted quantum error correction (EAQEC) codes can be constructed from previous known Hermitian self-orthogonal codes. Actually our method shows that previous constructed quantum MDS codes from Hermitian self-orthogonal codes can be transformed to MDS entanglement-assisted quantum codes with nonzero consumption parameter $c$ directly. We prove that MDS EAQEC $[[n, k, d, c]]_q$ codes with nonzero $c$ parameters and $d\leq \frac{n+2}{2}$ exist for arbitrary length $n$ satisfying $n \leq q^2+1$. Moreover any QECC constructed from $k$-dimensional Hermitian self-orthogonal codes can be transformed to $k$ different EAQEC codes. Moreover we prove that MDS entanglement-assisted quantum codes exist for all lengths $n\leq q^2+1$.\\

\end{abstract}

\section{Introduction}

The Hamming weight $wt_H({\bf a})$ of a vector ${\bf a} \in {\bf F}_q^n$ is the number of non-zero coordinate positions. The Hamming distance $d_H({\bf a}, {\bf b})$ between two vectors ${\bf a}$ and ${\bf b}$ is the Hamming weight $wt_H({\bf a}-{\bf b})$. For a code ${\bf C} \subset {\bf F}_q^n$, its Hamming distance $$d_H=\min_{{\bf a} \neq {\bf b}} \{d_H({\bf a}, {\bf b}),  {\bf a} \in {\bf C}, {\bf b} \in {\bf C} \},$$  is the minimum of Hamming distances $d_H({\bf a}, {\bf b})$ between any two different codewords ${\bf a}$ and ${\bf b}$ in ${\bf C}$. The shortening code of ${\bf C}$ at the $i$-th coordinate position is the subcode of ${\bf C}$ consisting of codewords in ${\bf C}$ whose $i$-th coordinate is zero. The minimum Hamming distance of  a linear code is its minimum Hamming weight.  For a linear $[n, k, d_H]_q$ code, the Singleton bound asserts $d_H \leq n-k+1$. When the equality holds, this code is an MDS code. The main conjecture of MDS codes claims that the length of an MDS code over ${\bf F}_q$ is at most $q+1$, except some trivial exceptional cases. In \cite{Ball} the main conjecture of MDS codes was proved for codes over prime fields. We refer to \cite{Lint,HP} for the theory of Hamming error-correcting codes.\\

 We say that two codes ${\bf C}_1$ and ${\bf C}_2$ in ${\bf F}_q^n$ are equivalent if ${\bf C}_2$ can be obtained from ${\bf C}_1$ by a permutation of coordinates and the multiplication of a Hamming weight $n$ vector ${\bf v}=(v_1, v_2, \ldots, v_n) \in {\bf F}_q^n$ on coordinates, where $v_i \neq 0$ for $i=1, \ldots, n$. That is $${\bf C}_2=\{{\bf c}=(c_1, \ldots c_n): (c_1, \ldots, c_n)=(v_1x_1, \ldots, v_nx_n), {\bf x} \in Perm({\bf C}_1)\},$$ where $Perm({\bf C}_1)$ is the code obtained from ${\bf C}_1$ by a coordinate permutation. Equivalent codes have the same coding parameters.\\

The Euclidean dual of a linear code ${\bf C}\subset {\bf F}_q^n$ is $${\bf C}^{\perp}=\{{\bf c}=(c_1, \ldots, c_n): \Sigma_{i=1}^n c_i x_i=0, \forall {\bf x}=(x_1, \ldots, x_n) \in {\bf C}\}.$$ The Hermitian dual of a linear code ${\bf C} \subset {\bf F}_{q^2}^n$ is $${\bf C}^{\perp_H}=\{{\bf c}=(c_1, \ldots, c_n): \Sigma_{i=1}^n c_i x_i^q=0, \forall {\bf x}=(x_1, \ldots, x_n) \in {\bf C}\}.$$  It is clear ${\bf C}^{\perp_H}=({\bf C}^{\perp})^q$, where $${\bf C}^q=\{(c_1^q, \ldots, c_n^q): (c_1 \ldots, c_n) \in {\bf C}\}.$$ The minimum distance of the Euclidean dual is called the dual distance and is denoted by $d^{\perp}$. The minimum distance of the Hermitian dual is the same as $d^{\perp}$. In general the linear code ${\bf C} \bigcap {\bf C}^{\perp}$ is called the Euclidean hull of the linear code ${\bf C} \subset {\bf F}_q$. The intersection ${\bf C} \bigcap {\bf C}^{\perp_H}$ is called the Hermitian hull of the linear code ${\bf C} \subset {\bf F}_{q^2}^n$.  \\

We refer to \cite{CPS,CS} and \cite{HP} Chapter 9 and \cite{Rains} for earlier results about self-dual codes over small fields and the theory of their classification.  It is always interesting in coding theory to construct self-dual MDS codes. There have been numerous papers on this topic, see \cite{GG08,Gulliver,JX17,Sok,Sok2,ZhangFeng}. However there is few constructions of Hermitian self-dual MDS codes, see \cite{Niu,GuoLi,Guo1}. For recent works on Hermitian self-orthogonal codes, we refer to \cite{Ball1,Ball2,Ball3} and a nice survey on quantum code, we refer to \cite{Ball4}. Obviously equivalent codes have different Hermitian dual codes as follows. Let ${\bf v}$ be a Hamming weight $n$ vector ${\bf v}=(v_1, v_2, \ldots, v_n) \in {\bf F}_{q^2}^n$, set ${\bf v} \cdot {\bf C}=\{(v_1c_1, \ldots, v_nc_n): \forall {\bf c}=(c_1, \ldots, c_n) \in {\bf C}\}$, where ${\bf C} \subset {\bf F}_{q^2}^n$ is a linear code,  then $$({\bf v} \cdot {\bf C})^{\perp_H}={\bf v}^{-q} \cdot {\bf C}^{\perp_H},$$ where ${\bf v}^{-q}=(v_1^{-q}, \ldots, v_n^{-q})$. The motivation of this paper is to use various Hamming weight $n$ vectors to control the dimension of the Hermitian hull.\\

Actually for a linear code ${\bf  C} \subset {\bf F}_{p^e}^n$ the following $l$-Galois dual ${\bf C}^{\perp_l}$, $0 \leq l \leq e-1$, can be defined $${\bf C}^{\perp_l}=\{{\bf c}=(c_1, \ldots, c_n): \Sigma_{i=1}^n c_i x_i^{p^l}=0, \forall {\bf x}=(x_1, \ldots, x_n) \in {\bf C}\},$$ see \cite{Fanzhang}. The intersection $Hull_l({\bf C})={\bf C} \bigcap {\bf C}^{\perp_l}$ is called the $l$-Galois hull of this code ${\bf C}$. Obviously when $e$ is even the Hermitian hull is $Hull_{\frac{e}{2}}({\bf C})$. The $l$-Galois dual of ${\bf v} \cdot {\bf C}$ is ${\bf v}^{-p^e} \cdot {\bf C}^{\perp_l}$. MDS codes with various $l$-Galois hull have been studied in \cite{MCao}.\\

Quantum error correction codes (QECCs) are necessary for quantum information processing and quantum computation. For constructions of quantum error correction codes we refer to \cite{Shor,Steane,CRSS} and \cite{HChen1,HChen2,HChen3,AKS,KS,Guar,JX,Ball1,KZ13,KZL14,CLZ15}. An QECC attaining the bound $$2d+k \leq n+2$$ is called MDS.  A lot of quantum MDS codes were constructed from Hermitian self-orthogonal codes, see \cite{Guar,JX,KZ13,KZL14,CLZ15,HCX}. Entanglement-assisted quantum error correction (EAQEC) codes were proposed in \cite{Brun,Fattal}. Comparing to a QECC an EAQEC code has one more parameter $c$ measuring the consumption of $c$ pre-shared copies of maximally  entangled states. From the basic results in \cite{Brun}, an EAQEC $[[n, k-h, d, n-k-h]]_q$ code and an EAQEC $[[n, n-k-h, d^{\perp}, k-h]]_q$ code can be obtained from a linear $[n, k, d]_q$ code with the $h$-dimension Euclidean hull, similarly from a linear $[n, k, d]_{q^2}$ code ${\bf C} \subset {\bf F}_{q^2}^n$ with the $h$-dimension Hermitian hull, an EAQEC $[[n, k-h, d, n-k-h]]_q$ code and an EAQEC $[[n, n-k-h, d^{\perp}, k-h]]_q$ can be constructed. The quantum Singleton bound asserts $$2d+k\leq n+c+2$$ for an EAQEC $[[n, k, d, c]]_q$ code when $d\leq \frac{n+2}{2}$, see \cite{Brun,GHW}.  In the above construction of EAQEC codes, the dimension $h$ of the Euclidean or Hermitian hull of a linear code is a key parameter to control the dimension and consumption parameter of an EAQEC code. This is an important motivation to construct equivalent linear codes with various hull dimensions, see \cite{LCC,GYHZ,FFLZ20,CZJL,CZJ,LES22,LEGL22}. An EAQEC code attaining this quantum Singleton bound is called an MDS EAQEC code. The construction of MDS QAECC code with large ranges of four parameters has been addressed in \cite{Kor,LCC,GYHZ,FFLZ20,CZJ,CZJL,GMCR,LES22,LEGL22} and references therein.\\

The construction of linear codes with $h$-dimension hull and their applications in EAQEC codes have been an active topic in recent years, we refer to \cite{LCC,GYHZ,Pellikaan,Sok}. It is interesting to know what coding parameter $[n, k, d]_q$ can be attainable by a linear code with the $h$-dimension hull.  On the other hand linear codes with one-dimension hull are efficient to determine their equivalence by the algorithm proposed in \cite{Sen1}. Some MDS codes with $h$-dimension Euclidean, Hermitian or $l$-Galois hull were constructed in \cite{LCC,FFLZ20,MCao} for some restricted code lengths. Then MDS entanglement-assisted quantum codes were constructed based on their codes with the $h$-dimension Euclidean or Hermitian hull. In this paper we show that this is actually not necessary. MDS entanglement-assisted quantum codes with nonzero $c$ parameters can be obtained from Hermitian self-orthogonal codes in \cite{Guar,JX,KZ13,KZL14,CLZ15,HCX,Ball1,Ball2,Ball3,FFLZ20} and \cite{CZJ,CZJL} directly.\\

In this paper we prove that for a nonnegative integer $h$ satisfying $0 \leq h \leq k$, a linear $[n, k]_{q^2}$ Hermitian self-orthogonal code is equivalent to a linear $h$-dimension Hermitian hull code. Then one QECC from Hermitian self-orthogonal code of the dimension $k$ can be transformed to $k$ different entanglement-assisted quantum codes with different $c$ parameters $1 \leq c\leq k$. Based on known MDS Hermitian self-orthogonal codes constructed in \cite{Guar,JX,KZ13,KZL14,CLZ15,HCX,Ball1,Ball2,Ball3,JX} and our Theorem 2.1 below, a lot of new MDS EAQEC codes with a large range of the $c$ parameter and various lengths, can be obtained directly from a linear MDS Hermitian self-orthogonal code. Our method indicates that all previous quantum MDS codes constructed from MDS Hermitian self-orthogonal codes can be transformed to MDS entanglement-assisted quantum codes with nonzero $c$ parameter.\\

\section{Arbitrary dimension Hermitian hull codes from Hermitian self-orthogonal codes}

Let ${\bf A}=(a_{ij})_{1 \leq i, j \leq n}$ be a $n \times n$ matrix with entries $a_{ij} \in {\bf F}_{q^2}$, then the matrix ${\bf \bar{A}}=(a_{ij}^q)_{1 \leq i, j \leq n}$ is the conjugate $n \times n$ matrix. ${\bf M}^{\tau}$ is always the transposition of a matrix ${\bf M}$.The following result is well-known.\\

{\bf Proposition 2.1.} {\em Let $({\bf I}_n, {\bf P})$ be a generator matrix of a linear self-dual $[2n, n]_q$ code ${\bf C} \subset {\bf F}_q^n$. Then ${\bf P} \cdot {\bf P}^{\tau}=-{\bf I}_n$. Therefore ${\bf P}$ is a non-singular $n \times n$ matrix. Let $({\bf I}_n, {\bf A})$ be a generator matrix of a linear Hermitian self-dual $[2n, n]_{q^2}$ code ${\bf C} \subset {\bf F}_{q^2}^n$. Then ${\bf A} \cdot {\bf \bar{A}}^{\tau}=-{\bf I}_n$. Therefore ${\bf A}$ is a non-singular $n \times n$ matrix.}\\

The main result of this Section is as follows.\\

{\bf Proposition 2.2.} {\em Let $k$ and $n$ be two positive integers satisfying $k \leq n$. Let $({\bf I}_k, {\bf P})$ be a generator matrix of a linear Hermitian self-orthogonal $[n, k]_{q^2}$ code ${\bf C} \subset {\bf F}_{q^2}^n$. Then ${\bf P} \cdot {\bf \bar{P}}^{\tau}=-{\bf I}_k$. Therefore ${\bf P}$ is a full rank $k \times (n-k)$ matrix.}\\

{\bf Proof.} One generator matrix of the Hermitian dual of ${\bf C}^{\perp_H}$ is of the form $(-{\bf \bar{P}}^{\tau}, {\bf I}_{n-k})$. Then we have a $k \times (n-k)$ matrix ${\bf Q}$ satisfying that $${\bf Q} \cdot (-{\bf \bar{P}}^{\tau}, {\bf I}_{n-k})=({\bf I}_k, {\bf P}).$$ Thus ${\bf Q}={\bf P}$ and ${\bf P} \cdot {\bf \bar{P}}^{\tau}=-{\bf I}_k$.\\

{\bf Theorem 2.1.} {\em Let $q \geq 3$ be a prime power.  If ${\bf C} \subset {\bf F}_{q^2}^n$ is  a linear Hermitian self-orthogonal $[n, k]_{q^2}$ code. Then there is a Hamming weight $n$ vector ${\bf v} \in {\bf F}_{q^2}$ such that ${\bf v} \cdot {\bf C}$ has the $h$-dimension Hermitian hull for nonnegative integer $h$ satisfying $0 \leq h \leq k$.}\\

{\bf Proof.} Let $({\bf I}_k, {\bf P})$ be a generator matrix of a linear self-dual $[n, k]_{q^2}$ code ${\bf C} \subset {\bf F}_{q^2}^n$. Since ${\bf P}$ is a full rank $k \times (n-k)$ matrix, we can rearrange the coordinate positions such that the first $k$ columns of ${\bf P}$ is a non-singular $k \times k$ matrix. Set ${\bf P}=({\bf P}_1, {\bf P}_2)$, where ${\bf P}_1$ is a $k \times k$ nonsingular matrix and ${\bf P}_2$ is a $k \times (n-2k)$ matrix.\\

Set ${\bf v}=(\lambda_1, \ldots, \lambda_{k-h}, 1, \ldots, 1)$, where $\lambda_1, \ldots, \lambda_{k-h}$ are $k-h$ nonzero elements in ${\bf F}_{q^2}$. We assume that all $\lambda_1^{q+1}, \ldots, \lambda_{k-h}^{q+1}$ are not $1$. This is possible since the field ${\bf F}_q$ has at least $3$ elements. Then a generator matrix of ${\bf v} \cdot {\bf C}$ is of the following form $({\bf D_{\lambda}}, {\bf P})$, where ${\bf D_{\lambda}}$ is a $k \times k$ non-singular matrix of the following form.\\

$$
\left(
\begin{array}{ccccccccccccc}
\lambda_1&0&0&\cdots&\cdots&\cdots&\cdots&0\\
0&\lambda_2&0&\cdots&\cdots&\cdots&\cdots&0\\
\cdots&\cdots&\cdots&\cdots&\cdots&\cdots&\cdots&\cdots\\
0&0&0&\cdots&\lambda_{k-h}&0&\cdots&0\\
0&0&0&\cdots&0&1&\cdots&0\\
0&0&0&\cdots&0&0&\cdots&1\\
\end{array}
\right)
$$
The Hermitian dual is ${\bf C}^{\perp_H}$ has a generator matrix of the form $(-{\bf \bar{P}}^{\tau}, {\bf I}_{n-k})$. Consider the following $(n-k) \times (n-k)$ nonsingular matrix ${\bf W}$,

$$
\left(
\begin{array}{cccc}
{\bf P}_1&{\bf P}_2\\
{\bf 0}_{n-2k, k}&{\bf I}_{n-2k}\\
\end{array}
\right)
$$

 , since ${\bf P}_1$ has the full-rank, ${\bf W} \cdot (-{\bf \bar{P}}^{\tau}, {\bf I}_{n-k})$ of the following form is also a generator matrix of ${\bf C}^{\perp_H}$. Here ${\bf 0}_{n-2k, k}$ is the $(n-2k) \times k$ zero matrix.\\

$$
\left(
\begin{array}{cccc}
{\bf I}_k&{\bf P}_1&{\bf P}_2\\
-{\bf \bar{P_2}}^{\tau}&{\bf 0}_{n-2k, k}&{\bf I}_{n-2k}\\
\end{array}
\right)
$$

Hence the Hermitian dual $({\bf v} \cdot {\bf C})^{\perp_H}={\bf v}^{-q} \cdot {\bf C}^{\perp_H}$ has one generator matrix ${\bf B}$ of the  following form.\\

$$
\left(
\begin{array}{cccc}
{\bf D}_{\lambda}^{-q}&{\bf P}_1&{\bf P}_2\\
-{\bf \bar{P_2}}^{\tau} \cdot {\bf D}_{\lambda}^{-q}&{\bf 0}_{n-2k, k}&{\bf I}_{n-2k}\\
\end{array}
\right)
$$

Then it is clear  that the last $h$ rows of the above generator matrix of ${\bf C}$ and the $k-h+1, k-h+2, \ldots, k$ rows of the above generator matrix  of $({\bf v} \cdot {\bf C})^{\perp_H}={\bf v}^{-q} \cdot {\bf C}^{\perp_H}$ are the same. Thus the dimension of the Hermitian hull $({\bf v} \cdot {\bf C}) \bigcap ({\bf v} \cdot {\bf C})^{\perp_H}$ is at least $h$.\\

 Let ${\bf C}_1$ be the $k-h$ dimension subcode of ${\bf v} \cdot {\bf C}$ generated by the first $k-h$ rows of the above generator matrix. We now prove that the dimension of the Hermitian hull is exactly $h$. Otherwise the natural mapping $$({\bf v} \cdot {\bf C}) \bigcap ({\bf v} \cdot {\bf C})^{\perp_H} \longrightarrow ({\bf v} \cdot {\bf C})/{\bf C}_1$$ is not injective and there is a nonzero ${\bf c} \in ({\bf v} \cdot {\bf C}) \bigcap ({\bf v} \cdot {\bf C})^{\perp_H}$ such that ${\bf c} \subset {\bf C}_1$.  Then we have a nonzero vector ${\bf x} \in {\bf F}_{q^2}^k$ such that its last $h$ coordinates are zero and a nonzero vector in ${\bf y}=({\bf y}_1, {\bf y}_2) \in {\bf F}_{q^2}^{n-k}$, satisfying $${\bf x} \cdot ({\bf D_{\lambda}}, {\bf P})={\bf y} \cdot {\bf B},$$ where ${\bf y}_1$ is a length $k$ vector and ${\bf y}_2$ is a length $n-2k$ vector. From the reading of the $k+1$ to $2k$ coordinates of this codeword, we have $${\bf x} \cdot {\bf P}_1={\bf y}_1 \cdot {\bf P}_1,$$ then ${\bf x}={\bf y}_1$ since ${\bf P}_1$ is nonsingular.  Notice that $${\bf y} \cdot {\bf B}$$ is of the form $({\bf y}_1 {\bf D_{\lambda}}^{-q}-{\bf y}_2 {\bf  \bar{P_2}}^{\tau} \cdot {\bf D_{\lambda}}^{-q}, {\bf y}_1 \cdot {\bf P}_1, {\bf y}_1 \cdot {\bf P}_2+{\bf y}_2)$. From the reading of the last $n-k$ coordinates of this codeword, we have ${\bf x} \cdot {\bf P}={\bf y}_1 \cdot {\bf P}+({\bf 0}_k, {\bf y}_2)$. Since ${\bf x}={\bf y}_1$, ${\bf y}_2={\bf 0}_{n-2k}$, where ${\bf 0}_l$ is a length $l$ zero vector.\\

Then this codeword $${\bf y} \cdot {\bf B}$$ is of the form $$({\bf y}_1 {\bf D_{\lambda}}^{-q}, {\bf y}_1 \cdot {\bf P}).$$
 Since the last $h$ coordinates of ${\bf x}$ are zero. Then the last $h$ coordinates of the length $k$ vector ${\bf y}_1$ are zero. From the reading of the first $k-h$ coordinates of this codeword, we have $\lambda_i^{q+1}=1$ for $i=1, \ldots, k-h$, since ${\bf x}={\bf y}_1$. This is a contradiction to the assumption that $\lambda_i^{q+1} \neq =1$. Then the conclusion is proved.\\

 The following result can be proved similarly.\\

{\bf Corollary 2.1.} {\em Let $q \geq 3$ be a prime power. Let ${\bf C} \subset {\bf F}_{p^e}$ be a linear $[n, k]_{p^e}$ code satisfying ${\bf C} \subset {\bf C}^{\perp_l}$. Then we have a $h$-dimension $l$-Galois hull equivalent code for nonnegative integer $h$ satisfying $0 \leq h \leq k$.}\\

From a similar argument as in the proof of Theorem 2.1 we have the following result.\\

{\bf Corollary 2.2.} {\em Let $q$ be a prime power satisfying $q \geq 3$. Let ${\bf C} \subset {\bf F}_{q^2}$ be a linear $[n, k]_{q^2}$ code with $l$-dimension Hermitian hull. Then for any nonnegative integer $l' \leq l$ there is a Hamming weight $n$ vector ${\bf v}$ such that ${\bf v} \cdot {\bf C}$ has $l' \leq l$ dimension Hermitian hull. Thus from an arbitrary linear $[n, k]_{q^2}$ code ${\bf C} \subset {\bf F}_{q^2}^n$ with its dual distance $d^{\perp} \leq \frac{n+2}{2}$ and $h$-dimension Hermitian hull, we have an EAQEC $[[n, n-k-l, d^{\perp}, k-l]]_q$ code, where $l$ is a nonnegative integer satisfying $l \leq h$.}\\

{\bf Proof.} We can assume that the generator matrix of ${\bf C}$ is of the following form.\\

$$
\left(
\begin{array}{ccccccc}
{\bf I}_l&{\bf P}_1&{\bf P}_2\\
{\bf  0}_{k-l, l}&{\bf I}_{k-l}&{\bf Q}\\
\end{array}
\right)
$$
such that the first $l$ rows, which is a $l \times n$ matrix, is a generator matrix of ${\bf C} \bigcap {\bf C}^{\perp_H}$. Then ${\bf P} \cdot {\bf \bar{P}}^{\tau}=-{\bf I}_l$, where ${\bf P}=({\bf P}_1, {\bf P}_2)$. The conclusion follows from a similar argument as the proof of Theorem 2.1.\\

We recall some nice Hermitian self-orthogonal codes constructed recently by S. Ball and R. Vilar.\\

{\bf Theorem 2.2 (see Theorem 4 in \cite{Ball1})} {\em Let $k$ be a positive integer satisfying $k \leq q$ and $k \neq q-1$. Then there is a Hermitian self-orthogonal $[q^2+1, k, q^2+2-k]_{q^2}$ MDS codes.}\\

{\bf Theorem 2.3 (see Theorem 5.2 in \cite{Ball2}).} {\em Let $q=2^r$, $ r \geq 3$ is odd. Then there is a Hermitian self-orthogonal $[q^2+1, q-1, q^2-q+3]_{q^2}$ code.}\\

{\bf Theorem 2.4 (see Theorem 3.4 in \cite{Ball2}).} {\em Let $q$ be a prime power, $k$ and $n$ be two positive integers satisfying $n \leq q^2$ and $k \leq q-1$. There is a linear $[n, k, n-k+1]_{q^2}$ Hermitian self-orthogonal MDS code if and only if there is a polynomial $g(x) \in {\bf F}_{q^2}[x]$ of degree at most $(q-k)q-1$, such that $g(x)+g(x)^q$ has $q^2-n$ distinct roots when evaluated at $x \in {\bf F}_{q^2}$.}\\

For example, let $q$ be a prime power and $t$ be a divisor of $q+1$ and $f(x) \in {\bf F}_{q^2}[x]$ be a polynomial such that $1+(q+1)\deg (f) \leq (q-k)q-1$. Let $M$ be the number of distinct roots of $f(x^{q+1})$ in ${\bf F}_{q^2}$. Set $N=1+t(q-1)+M$. Then there is an MDS Hermitian self-orthogonal $[q^2-N, k, q^2-N+1-k]_{q^2}$ code, see Example 3.5 in \cite{Ball}. When $q$ is an odd prime power, $t$ is a divisor of $\frac{q+1}{2}$, for any given degree $\deg(f)$ satisfying $1+(q+1)\deg(f) \leq (q-k)q-1$, there is an MDS Hermitian self-orthogonal $[q^2-1-t(q-1)-\deg(f)(q+1), k, q^2-t(q-1)-\deg(f)(q+1)-k]_{q^2}$ code.\\

From Theorem 2.1, and previous results due to S. Ball and R. Vilar in \cite{Ball1,Ball2} we have the following result Corollary 2.1.\\

{\bf Corollary 2.3.} {\em 1) Let $q$ be a prime power satisfying $q \geq 3$, $k$ be a positive integer satisfying $k\leq q$ and $k \neq q-1$ and $h$ be a nonnegative integer satisfying $0 \leq h \leq k$. Then we have a $h$-dimension Hermitian hull $[q^2+1, k, q^2+2-k]_{q^2}$ code.\\

2) Let $q=2^r$, $ r \geq 3$ is odd, and $h$ be a nonnegative integer satisfying $0 \leq h \leq q-1$, we have a $h$-dimension Hermitian hull $[q^2+1, q-1, q^2-q+2]_{q^2}$ code.}\\

\section{New MDS  EAQEC codes}

We present applications to construct MDS EAQEC codes with various four parameters. The following result is direct from Theorem 2.1.\\

{\bf Corollary 3.1.} {\em Let $q$ be a prime power satisfying $q \geq 3$ and $h$ be a positive integer satisfying $0 \leq h \leq k$. If there is a Hermitian self-orthogonal $[n, k, d]_{q^2}$ code with its dual distance $d^{\perp}$, then we can construct an EAQEC $[[n, n-k-h, d^{\perp}, k-h]]_q$ code.}\\

From the optimal Hermitian self-dual codes constructed in \cite{Sok2} we give tables of various length EAQEC codes over small fields in \cite{HChen22}. From these $h$-dimension Hermitian hull codes in Corollary 2.1 based on Ball-Vilar Hermitian self-orthogonal codes, we have the following result immediately. \\

{\bf Corollary 3.2.} {\em 1) Let $q$ be a prime power satisfying $q \geq 3$, $k$ and $h$ be one positive integer and one nonnegative integer satisfying $k \leq q$ and $k \neq q-1$ and $0 \leq h < k$. Then we can construct an MDS EAQEC $[[q^2+1, q^2+1-k-h, k+1, k-h]]_q$ code.\\

2) Let $q=2^r$, $r \geq 3$ is odd, and $n$ be a nonnegative integer satisfying $0\leq h \leq q-1$. Then we can construct an MDS EAQEC $[[q^2+1, q^2-q+2-h, q, q-1-h]]_q$ code.}\\

The following result follows from Ball-Vilar Theorem 3.4 and Example 3.5 in \cite{Ball2},  and our Theorem 2.1  immediately.\\

{\bf Corollary 3.3.} {\em 1) Let $q$ be a prime power satisfying $q \geq 3$, $k$ and $n$ be two positive integers satisfying $n \leq q^2$ and $k \leq q-1$ and $h$ be a nonnegative integer satisfying $0 \leq h \leq k$. Suppose that there is a polynomial $g(x) \in {\bf F}_{q^2}[x]$ of degree at most $(q-k)q-1$, such that $g(x)+g(x)^q$ has $q^2-n$ distinct roots when evaluated at $x \in {\bf F}_{q^2}$. Then there is an MDS EAQEC $[[n, n-k-h, k+1, k-h]]_q$ code.\\

2) Let $q$ be a prime power satisfying $q \geq 3$, $k$ be positive integer satisfying $k \leq q-1$, $t$ be a divisor of $\frac{q+1}{2}$, $u$ be a positive integer satisfying $1+(q+1)u \leq (q-k)q-1$ and $h$ be a nonnegative integer satisfying $0 \leq h \leq k$. Then there is an MDS EAQEC $[[q^2-1-t(q-1)-u(q+1), q^2-1-t(q-1)-u(q+1)-k-h, k+1, k-h]]_q$ code.}\\

A lot of Hermitian self-orthogonal codes were constructed in our previous paper \cite{HCX} therefore a lot of MDS EAQEC codes can be obtained from MDS Hermitian self-orthogonal codes in \cite{HCX}. Let $q$ be a prime power, $m=2k+1$ be an odd divisor of $q+1$, $w$ be positive integer satisfying $w < \frac{(k+1)(q-1)}{2k+1}$. Then an MDS Hermitian self-orthogonal $[\frac{q^2-1}{m}, w, \frac{q^2-1}{m}-w+1]_{q^2}$ code was constructed in Theorem 2.1 of \cite{HCX}. Then we have the following result from Theorem 2.1.\\

{\bf Corollary 3.4.} {\em Let $q$ be a prime power satisfying $q \geq 3$, $m=2k+1$ be an odd divisor of $q+1$, $w$ be positive integer satisfying $w < \frac{(k+1)(q-1)}{2k+1}$, and $h$ be a nonnegative integer satisfying $0 \leq h \leq w$. Then an MDS EAQEC $[[\frac{q^2-1}{m}, \frac{q^2-1}{m}-w-h, w+1, w-h]]_q$ code can be constructed.}\\

Set $q+1=\lambda r$, where $r=2k+1$ is odd. Let $w$ be a positive integer satisfying $w \leq \frac{q-1}{2}$. Then an MDS Hermitian self-orthogonal $[\lambda(q-1), w, \lambda(q-1)-w+1]_{q^2}$ code is constructed from Theorem 2.1 in \cite{HCX}. We have an MDS EAQEC  $[[\lambda (q-1), \lambda(q-1)-w-h, w+1, w-h]]_q$ code. Many MDS QAEQC codes constructed in \cite{FFLZ20} can be recovered from this construction.\\

Let $q$ be an odd prime power. Suppose that $m_1$ and $m_2$ are odd divisors of $q+1$ satisfying $\gcd(m_1, m_2)=1$. Let $k$ be a positive integer satisfying $1 \leq k \leq \frac{q-1}{2}$. Then an MDS Hermitian self-orthogonal $[\frac{q^2-1}{m_1}+\frac{q^2-1}{m_2}-\frac{q^2-1}{m_1m_2}, k, \frac{q^2-1}{m_1}+\frac{q^2-1}{m_2}-\frac{q^2-1}{m_1m_2}-k+1]_{q^2}$ code was constructed in Theorem 2.3 in \cite{HCX}. We have the following result from Theorem 2.1.\\

{\bf Corollary 3.5.} {\em Let $q$ be a prime power satisfying $q \geq 3$. Suppose that $m_1$ and $m_2$ are odd divisors of $q+1$ satisfying $\gcd(m_1, m_2)=1$. Let $k$ be a positive integer satisfying $1 \leq k \leq \frac{q-1}{2}$ and $h$ be a nonnegative integer satisfying $0 \leq h \leq k$. We can construct an MDS EAQEC $[[\frac{q^2-1}{m_1}+\frac{q^2-1}{m_2}-\frac{q^2-1}{m_1m_2}, \frac{q^2-1}{m_1}+\frac{q^2-1}{m_2}-\frac{q^2-1}{m_1m_2}-k-h, k+1, k-h]]_q$ code.}\\

Let $q=2^ha+1$ be an odd prime power, where $a$ is odd, if $m=2^{h_1}a_1 \geq 6$ is an even divisor of $q-1$, that is, $h_1 \leq h$ and $a_1$ is an odd divisor of $a$. Let $k$ be a positive integer satisfying $1 \leq k \leq \frac{q+1}{2}+2^{h-h_1}\frac{a}{a_1}-1$. Then a linear Hermitian self-orthogonal $[\frac{q^2-1}{m}, k, \frac{q^2-1}{m}-k+1]_{q^2}$ code was constructed in \cite{CLZ15} Theorem 4.11 and Theorem 3.1 of \cite{HCX}. Then we have the following MDS EAQEC codes.\\

{\bf Corollary 3.6.} {\em Let $q=2^ha+1$ be an odd prime powe satisfying $q \geq 3$, where $a$ is odd, if $m=2^{h_1}a_1 \geq 6$ is an even divisor of $q-1$, that is, $h_1 \leq h$ and $a_1$ is an odd divisor of $a$. Let $k$ be a positive integer satisfying $1 \leq k \leq \frac{q+1}{2}+2^{h-h_1}\frac{a}{a_1}-1$ and $h$ be a nonnegative integer satisfying $0 \leq h \leq k$. Then we can construct an MDS EAQEC $[[\frac{q^2-1}{m}, \frac{q^2-1}{m}-k-h, k+1, k-h]]_q$ code.}\\

In the following table we summarize some MDS EAQEC codes from Hermitian self-orthogonal codes constructed by Ball-Vilar in \cite{Ball1,Ball2} and previous known Hermitian self-orthogonal codes constructed in \cite{JX,KZL14,CLZ15,HCX}. Certainly we do not list all MDS EAQEC codes from previous known Hermitian self-orthogonal codes in \cite{Guar,JX,KZ13,KZL14,CLZ15,HCX,Ball2,Ball3,HCX,CZJL,CZJ}.  Actually for a linear $h$-dimension Hermitian hull code we can get at least $h$ entanglement-assisted quantum codes from Corollary 2.2. Therefore for an arbitrary length generalized Reed-Solomon codes over ${\bf F}_{q^2}$ with its dual distance $d^{\perp} \leq \frac{n+2}{2}$, if the dimension $h$ of its Hermitian hull is determined, we get $h$ MDS EAQEC codes. Then a lot of new MDS EAQEC codes can be obtained.\\

{\bf Corollary 3.7.} {\em Let $q$ be a prime power satisfying $q \geq 3$. Let ${\bf C} \subset {\bf F}_{q^2}^n$ be a generalized Reed-Solomon code with $h$-dimension Hermitian hull and the dual distance $d^{\perp}=k+1 \leq \frac{n+2}{2}$. Then we can construct an MDS EAQEC $[[n, n-k-l, k+1, k-l]]_q$ code, where $l$ is a nonnegative integer satisfying $0 \leq l \leq h$.}\\

From Corollary 3.7  for an arbitrary length Reed-Solomon $[n, k, n-k+1]_{q^2}$ code ${\bf C} \subset {\bf F}_{q^2}^n$, we need to determine the dimension $h$ of its Hermitian hull. If $ h \neq k$ and $h \neq n-K$, that is, ${\bf C}$ is not Hermitian self-orthogonal $${\bf C} \subset {\bf C}^{\perp_H}$$ and not Hermitian dual-containing $${\bf C}^{\perp_H} \subset {\bf C},$$ then $h+1$ different MDS EAQEC codes with $k+1 \leq \frac{n+2}{2}$ or $n-k+1 \leq \frac{n+2}{2}$ can be obtained. The $c$ parameters of these $h+1$ MDS EAQEC codes are different.\\

{\bf Corollary 3.8.} {\em Let $q$ be a prime power satisfying $q \geq 3$. For any given length  $n \leq q^2+1$, there exists at least one MDS EAQEC $[[n, d, k, c]]_q$ code with nonzero $c$ parameter.}\\

{\bf Proof.} The case of length $q^2+1$ is proved in Corollary 3.2. Then we take an arbitrary Reed-Solomon $[n, k, n-k+1]_{q^2}$ code ${\bf C} \subset {\bf F}_{q^2}$ with the dimension $k \leq \frac{n}{2}$. If the dimension of the Hermitian hull of this code is $k$, that is, ${\bf C}$ is Hermitian self-orthogonal. The conclusion follows from Theorem 2.1. If $h<k$. The conclusion follows from Corollary 3.7.\\

Since MDS entanglement-assisted quantum codes with nonzero $c$ parameters constructed in all previous papers \cite{Kor,GMCR,CZJL,CZJ,LCC,FFLZ20,GYHZ,MCao,Sok} are for special lengths, the above result is much stronger than previous results. In many computational constructions of MDS entanglement-assisted quantum codes in \cite{Kor,GMCR,CZJL,CZJ,LCC,FFLZ20,GYHZ,MCao,Sok} for some special lengths $n \leq q^2+1$, there are indeed Hermitian self-orthogonal codes for these lengths. Therefore the construction of MDS generalized Reed-Solomon codes with $h < \min\{k, n-k\}$ dimensional Hermitian hull can follow from our Theorem 2.1 immediately.\\

\begin{table}[hb]
\centering
\caption{Parameters of MDS EAQEC codes over ${\bf F}_q$} \label{table-bounds-s-2}
\begin{tabular}{|c|l|l|l|l|l|l|}\hline
$n$&$k$&$d\leq \frac{n+2}{2}$&$c$&Conditions\\ \hline
$q^2+1$&\tabincell{l}{$q^2+1$\\$-k-h$}&$k+1$&$k-h$&\tabincell{l}{$0 \leq h \leq k \leq q$, \\$k \neq q-1$}\\ \hline
$q^2+1$&$q^2+2-q-h$&$q$&$q-1-h$&\tabincell{l}{$q=2^r$, $r\geq 3$ \\odd, $0\leq h \leq q-1$}\\ \hline
\tabincell{l}{$q^2-1-$\\$t(q-1)-$\\$u(q+1)$}&\tabincell{l}{$q^2-1-$\\$t(q-1)-$\\$u(q+$\\$1)-k-h$}&$k+1$&$k-h$& \tabincell{l}{$q$ odd, $k \leq q-1,$\\ $ t|\frac{q+1}{2}, 1+u(q+1)$\\$\leq (q-k)q-1,$\\$0\leq h \leq k$} \\ \hline
$q^2-s$&\tabincell{l}{$q^2-s$\\$-k-h$}&$k+1$&$k-h$&\tabincell{l}{$0 \leq h \leq k$, \\$0 \leq s \leq \frac{q}{2}-1$\\$\frac{q}{2} \leq k \leq q-s-1$}\\ \hline
$\frac{q^2+1}{5}$&$\frac{q^2+1}{5}-k-h$&$k+1$&$k-h$&\tabincell{l}{$q=20m+3$, or \\$q=20m+7$\\ $1 \leq k \leq \frac{q+3}{2}$\\ $0 \leq h \leq k$} \\ \hline
$2t(q-1)$&$2t(q-1)-k-h$&$k+1$&$k-h$&\tabincell{l}{ $8|(q+1)$, \\$t|(q+1)$, \\$t$ odd, \\$1\leq k \leq 6t-2$, \\$0 \leq h \leq k$} \\ \hline
\tabincell{l}{$\frac{q^2-1}{m_1}$\\$+\frac{q^2-1}{m_2}-$\\$\frac{q^2-1}{m_1m_2}$}&\tabincell{l}{$\frac{q^2-1}{m_1}+$\\$\frac{q^2-1}{m_2}-$
\\$\frac{q^2-1}{m_1m_2}$\\$-k-h$}&$k+1$&$k-h$&\tabincell{l}{$m_1|(q+1)$,\\ $m_2|(q+1)$, \\$\gcd(m_1, m_2)=1$, \\$k \leq \frac{q-1}{2}$} \\ \hline
$\frac{q^2-1}{m}$&$\frac{q^2-1}{m}-k-h$&$k+1$&$k-h$&\tabincell{l}{$q=2^ha+1$, \\$m=2^{h_1}a_1$, \\$m|q-1$, $a$ odd, \\$a_1$ odd, $a_1|a$, \\$1\leq k \leq \frac{q+1}{2}+$\\$2^{h-h_1}\frac{a}{a_1}-1$, \\$0 \leq h \leq k$} \\ \hline
$n$&$n-k-l$&$k+1$&$k-l$&\tabincell{l}{$n\leq q^2+1$, \\$h=\dim_{{\bf F}_{q^2}}({\bf C} \bigcap {\bf C}^{\perp_H})$, \\$l \leq h$} \\ \hline
\end{tabular}
\end{table}

\section{Conclusion}
We proved that there is an equivalent linear $h$-dimension Hermitian hull code for each Hermitian self-orthogonal $[n, k]_{q^2}$ code if $0 \leq h \leq k$. Then an MDS QECCs can be transformed to several MDS EAQEC codes with different $c$ parameters. They were obtained from previous known Hermitian self-orthogonal codes directly. Comparing with previous constructions in  \cite{LCC,GYHZ,FFLZ20,CZJ,CZJL} a lot of new MDS EAQEC codes have been presented in this paper. In particular arbitrary length $n \leq q^2+1$ nonzero $c$ parameter MDS EAQEC codes over ${\bf F}_q$ exist. Therefore there are much more MDS EAQEC codes with nonzero $c$ parameters than MDS QECCs. It was also proved that MDS entanglement-assisted quantum code exists for each length $n \leq q^2+1$, this is much stronger than previous construction of such optimal codes for special lengths.\\


\begin{thebibliography}{10}

\bibitem{AKS} S. A. Aly, A. Klappenecker and P. K. Sarvepalli, On quantum and classical BCH codes, IEEE Trans. Inf. Theory, vol. 53, no. 3, pp. 1183-1188, 2007.

\bibitem{Ball} S. Ball, On large subsets of a finite vector space in which every subset of basis size is a basis, Journal of the European Mathematical Society, vol. 14, pp. 733-748, 2012.

\bibitem{Ball1} S. Ball, Some constructions of quantum MDS codes, Des., Codes and Cryptogra. vol. 89, pp. 811-821, 2021.

\bibitem{Ball2} S. Ball and R. Vilar, Determining when a truncated generalised Reed-Solomon code is Hermitian self-orthogonal, IEEE Trans. Inf. Theory, vol. 68, pp. 3796-3805,  2022.

\bibitem{Ball3} S. Ball and R. Vilar, The geometry of Hermitian orthogonal codes, Journal of Geometry, vol. 113, artical no. 7, 2022.


\bibitem{Ball4} S. Ball, A. Centelles and F. Huber, Quantum error-correcting codes and their geometries, Annale de l'institut Henri Poincare, to appear, 2022.


\bibitem{Brun} T. A. Brun, I. Devetak and M-H. Hsieh, Correcting quantum errors with entanglemnent, Science, vol. 304 (5798), pp. 436-439, 2006.

\bibitem{CRSS} A. Calderbank, E. M. Rains, P. W. Shor and N. J. A. Sloane, Quantum error correction via codes over $GF(4)$, IEEE Trans. Inf. Theory, vol. 44, no. 4, pp. 1369-1387, 1998.

\bibitem{MCao} M. Cao, MDS codes with Galois hulls of arbitrary dimensions and the related entanglement-assisted quantum error correction, IEEE Trans. Inf. Theory, vol. 67, no. 12, pp. 7964-7984, 2021.

\bibitem{CLZ15} B. Chen, S. Ling and G. Zhang, Applications of constacyclic codes to quantum MDS codes, IEEE Trans. Inf. Theory, vol. 61, no. 3, pp. 1474-1483, 2015.

\bibitem{HChen1} H. Chen, Some good quantum error-correcting codes from algebraic geometric codes, IEEE Trans. Inf. Theory, vol. 47, no. 5, pp. 2059-2061, 2001.


\bibitem{HChen2} H. Chen, C. Xing and S. Ling, Asymptotically good quantum codes exceeding the Ashikhmin-Litsyn-Tsafasman bound,  IEEE Trans. Inf. Theory, vol. 47, no. 5, pp. 2055-2058, 2001.


\bibitem{HChen3} H. Chen, C. Xing and S. Ling, Quantum codes from concatenated algebraic geometric codes, IEEE Trans. Inf. Theory, vol. 51, no. 8, pp. 2915-2920, 2005.



\bibitem{HChen22} H. Chen, On hull-variation problem of equivalent linear codes, arXiv:2206.14516, 2022.


\bibitem{CZJL} X. Chen, S. Zhu, W. Jiang and G. Luo, A new family of EAQMDS codes constructed from constacyclic codes, Des., Codes and Cryptogra., vol. 89, pp. 2179-2193, 2021.

\bibitem{CZJ} X. Chen, S. Zhu and W. Jiang, Cyclic codes and some new entanglement-assisted quantum MDS codes, Des., Codes and Cryptogra., vol. 89, pp. 2533-2551, 2021.

\bibitem{CPS} J. H. Conway, V. Pless and N. J. A. Sloane, Self-dual codes over $GF(3)$ and $GF(4)$ of length not exceeding $16$, IEEE Trans. Inf. Theory, vol. 25, no. 3, pp. 312-322, 1979.

\bibitem{CS} J. H. Conway and N. J. A. Sloane, A new upper bound on the minimal distance of self-dual codes. IEEE Trans. Inf. Theory vol. 36, pp. 1319-1333, 1990.

\bibitem{FFLZ20} W. Fang, F. Fu, L. Li and S. Zhu, Euclid and Hermitian hulls of MDS codes and their application to quantum codes, IEEE Trans. Inf. Theory vol. 66. no. 6, pp. 3527-3537, 2020.

\bibitem{Fanzhang} Y. Fan and L. Zhang, Galois self-dual constacyclic codes, Des., Codes and Cryptogra., vol. 84, no.3 pp. 473-492, 2017.

\bibitem{Fattal} D. Fattal, T. S. Cubitt, Y. Yamamoto, S. Bravyi and I. L. Chuang, Entanglement in the stablizer formalism, arXiv:quant-ph/04-6168, 2004.

\bibitem{GYHZ} Y. Gao, Q. Yue, X. Huang and J. Zeng, Hulls of generalized Reed-Solomon codes via Goppa codes and their applications to quamtum codes, IEEE Trans. Inf. Theory, vol. 67, no. 10, pp. 6619-6626, 2021.

\bibitem{GMCR} C. Galindo, F. Hernando and D. Ruano, Entanglement-assisted quantum codes from RS codes and BCH codes with extension degree two, Quantum Inf. Process., vol. 20, pp. 158, 2016.

\bibitem{GG08} M. Grassl and T. A. Gulliver, On self-dual MDS codes, Proc. Int. Symp. Inf. Theory, pp. 1954-1957, 2008.

\bibitem{GHW} M. Grassl, F. Huber and A, Winiter, Entropic proofs of Singleton bounds for quantum error-correcting codes, IEEE Trans. Inf. Theory, vol. 68, no. 6, pp. 3942-3950, 2021.

\bibitem{Guar} G. G. La Guardia, New quantum MDS codes, IEEE Trans. Inf. Theory, vol. 57, no. 8, pp. 5551-5554, 2011.

\bibitem{GuoLi} G. Guo and R. Li, Hermitian self-dual GRS and entended GRS codes, IEEE Commun. Lett., vol. 25, no. 4, pp. 1062-1065, 2021.

\bibitem{Guo1} G. Guo, R. Li, Y. Liu and H. Song, Duality of generalized twisted Reed-Solomon codes and Hermitian self-dual MDS and NMDS codes, arXiv:2202.11457, 2022.

\bibitem{Gulliver} T. A. Gulliver, J-L. Kim and Y. Lee, New MDS or near MDS codes, IEEE Trans. Inf. Theory, vol. 54, no. 9, pp. 4354-4360, 2008.


\bibitem{HCX} X. He, L. Xu and H. Chen, New $q$-ary quamtum MDS codes with distances bigger than $\frac{q}{2}$, Quantum Inf. Process., vol. 15, pp. 2745-2758, 2016.

\bibitem{HP} W. C. Huffman and V. Pless, Fundamentals of error-correcting codes, Cambridge University Press, Cambridge, U. K., 2003.

\bibitem{JX} L. Jin and C. Xing, Euclid and Hermitian self-orthogonal algebraic geometric codes and their applications to quantum codes, IEEE Trans. Inf. Theory, vol. 58, no. 8, pp. 5484-5489, 2012.

\bibitem{JX17}  L. Jin and C. Xing, New MDS self-dual codes from generalized Reed-Solomon codes,  IEEE Trans. Inf. Theory, vol. 63, no. 3, pp. 14-34-1438, 2017.







\bibitem{LCC} G. Luo, X. Cao and X. Chen, MDS codes with hulls of arbitray dimensions and their quantum error correction, IEEE Trans. Inf. Theory, vol. 65, no. 5, pp. 2944-2952, 2019.

\bibitem{LES22} G. Luo, M. F. Ezerman and S. Ling, Entanglement-assisted and subsystem quantum codes: new propagation rule and construction, arXiv:2206.09782, 2022.

\bibitem{LEGL22} G. Luo, M. F. Ezerman, M. Grassl and S. Ling, How much entanglement does a quantum code need? Preprint, 2022.

\bibitem{Lint} J. H. van Lint, Introduction to the coding theory, GTM 86, Third and Expanded Edition, Springer, Berlin, 1999.

\bibitem{KZ13} X. Kai and S. Zhu, New quantum MDS codes from negacyclic codes, IEEE Trans. Inf. Theory, vol. 59, no. 2, pp. 1193-1197, 2013.


\bibitem{KZL14} X. Kai, S. Zhu and P. Li, Constacyclic codes and some new quantum MDS codes, IEEE Trans. Inf. Theory, vol. 60, no. 4, pp. 2080-2086, 2014.

\bibitem{KS} A. Klappenecker and P. K. Sarvepalli, Clifford code constructions of operators quantum error correcting codes, IEEE Trans. Inf. Theory, vol. 54, no. 12, pp. 5760-5765, 2008.

\bibitem{Kor} M. E. Koroglu, New entanglement-assisted MDS quantum codes from constacyclic codes, Quantum Inf. Process., vol. 18, pp. 1-18, 2018.

\bibitem{Niu} Y. Niu, Q. Yue, Y. Wu and L. Hu, Hermitian self-dual, MDS and generalized Reed-Solomon codes, IEEE Commun. Lett., vol. 23, no. 5, pp. 781-784, 2019.




\bibitem{Pellikaan} F. R. F. Pereira, R. Pellikaan, G. G. La Guardia and F. Marcos, Entanglement-assisted quantum codes from algebraic geometry codes, IEEE Trans. Inf. Theory, vol. 67, no. 11, pp. 7110-7120, 2021.

\bibitem{Rains} E. M. Rains and N. J. A. Sloane, Self-dual codes, In "Handbook of Coding Theory", eds, V. Pless and W. C. Huffman, pp. 177-294, Elsevier, Amsterdam, 1998.


\bibitem{Sen1} N. Sendrier, Finding the permutation between equivalent linear codes, IEEE Trans. Inf. Theory, vol. 46, no. 4, pp. 1193-1203, 2000.

\bibitem{Shor} P. W. Shor, Scheme for redcuing decoherence in quantum memory, Phys. Rev. A, vol. 52, pp. R2493-2496, 1995.



\bibitem{Sok} Lin Sok, Explicit constructions of MDS slef-dual codes, IEEE Trans. Inf. Theory, vol. 66, no. 6, pp. 3603-3615, 2020.


\bibitem{Sok2} Lin Sok, Construction of optimal Hermitian self-dual codes from unitary matrices, arXiv:1911.10456, 2019.

\bibitem{Steane} A. M. Steane, Mutiple particle interference and quantum error correction, Proc. Roy. Soc. London, vol. 452, pp. 2551-2577, 1996.




\bibitem{ZhangFeng} A. Zhang and K. Feng, A unified approch to construct MDS self-dual codes via Reed-Soloon code, IEEE Trans. Inf. Theory, vol. 66, no. 6, pp. 3650-3656, 2020.




\end{thebibliography}
\end{document}